\documentclass[letterpaper]{article}
\usepackage{amsmath}
\usepackage{graphicx}
\title{Environmental benefits of enhanced surveillance technology on airport departure operations}
\author{Pierrick Burgain and Eric Feron~\thanks{Burgain is with Metron Aviation, Inc, {\tt pburgain@gmail.com}. Feron is with the Georgia Institute of Technology, {\tt feron@gatech.edu}. This work was funded by Thales Air Operations.}}

\begin{document}
\bibliographystyle{IEEEtran}
\maketitle

Airport departure operations constitute an important source of airline delays and passenger frustration. Excessive surface traffic is the cause of increased controller and pilot workload; It is also the source of increased emissions; It worsens traffic safety and often does not yield improved runway throughput. Acknowledging this fact, this paper explores some of the feedback mechanisms by which airport traffic can be optimized in real time according to its current degree of congestion. In particular, it examines the environmnetal benefits that improved surveillance technologies can bring in the context of gate- or spot-release aircraft strategies.
It is shown that improvements can lead yield 4\% to 6\% emission reductions for busy airports like New-York La Guardia or Seattle Tacoma. These benefits come on top of the benefits already obtained by adopting threshold strategies currently under evaluation.

\section{Introduction: Lean airport operations mean fewer aircraft on the taxiways}

\subsection{Air traffic growth and airport congestion}
The U.S. National Airspace System (NAS) is expected to grow about 2.4\% per year over the next 20 years and accommodate around 1.6 times today's traffic level by 2028 \cite{greenair2010,nextGen,Ky2006,Arbuckle2006}. The anticipated growth of air traffic is expected to bring additional concerns to an already congested system  \cite{Joint2004, Thanh2006}. In particular, airports constitute one of the major obstacles to the growth of air traffic.
Even though the development
of smaller regional airports is expected, it is predicted that major airports will keep running at full capacity \cite{nextGen}. In some cases, airports will not be able to expand their capacity sufficiently to
meet the increasing demand. Airports like New-York LaGuardia
will be physically restrained by the lack of space for new runways or
ramps. Other airports will not be able to grow physically because of significant opposition from local communities. 

\subsection{Environmental impact of airports}
The contribution of aviation to CO$_{2}$ and NOx emissions around airports is expected to increase
 significantly by 2025 and beyond \cite{U.S.1999,Intergovernmental1999}. Hence, environmental impacts
 are expected to be the fundamental constraint on air transportation growth \cite{Waitz2004}.
  Indeed, concerns over pollution have forced governmental, environmental, and regulatory agencies to
   start implementing emissions abatement procedures at certain airports, such as La Guardia \cite{Boeing}.
   Starting in 2012, in the European Union, CO$_{2}$ emissions will be capped at the average 2004$/$2006 levels.
   This will concern all flights arriving at, and departing from, European airports.
   In the United States, Section 231 of the Clean Air Act gives the Environmental Protection Agency (EPA)
   the authority to regulate aircraft emissions, and to adopt emissions standards for U.S.-flagged aircraft \cite{United2003}.
   Additionally, many efforts are being conducted towards emissions-reduction technologies and concepts, such as electric taxi, with new operational
   procedures expected to provide the greatest near-term benefits \cite{Waitz2004}. On the ground, the level of environmental nuisance can be directly tied to the number of aircraft whose engines are running at any given time. These aircraft are typically those in the taxi phase. Thus, the fewer aircraft taxiing on the airport surface at any time, the lower the environmental impact of the airport.

\subsection{Current initiatives for improving departure operations}
To tackle this environmental issue, the NextGen
concept of operations \cite{nextGen} encourages research in surface
traffic operations aimed at lowering emissions and improving surface
traffic planning.
Likewise, according to \cite{dman}, EUROCONTROL is currently
fine-tuning the Airport Collaborative Decision Making Departure
Manager (CDM DMAN) concept of operations and is preparing the
necessary implementation guidelines. 
DMAN incorporates
Collaborative Decision-Making (CDM) as a tool for managing departure
operations. DMAN ``keeps the number of aircraft on the taxiway at
an optimal level'' and ``keeps the taxiways open for other traffic without
blocking stands for arrivals, reduces controller workload,
improves punctuality and predictability, facilitates co-operation between
aerodrome ATC, airlines and airport operators, enhances
CFMU [i.e. Central Flow Management Unit slot-revisions] and slot compliance,
and exploits the departure capacity of the respective
runway'' \cite{dman}. 
\subsection{Analytic research efforts}
\begin{figure}[ht]
\centering
\includegraphics[scale = .30]{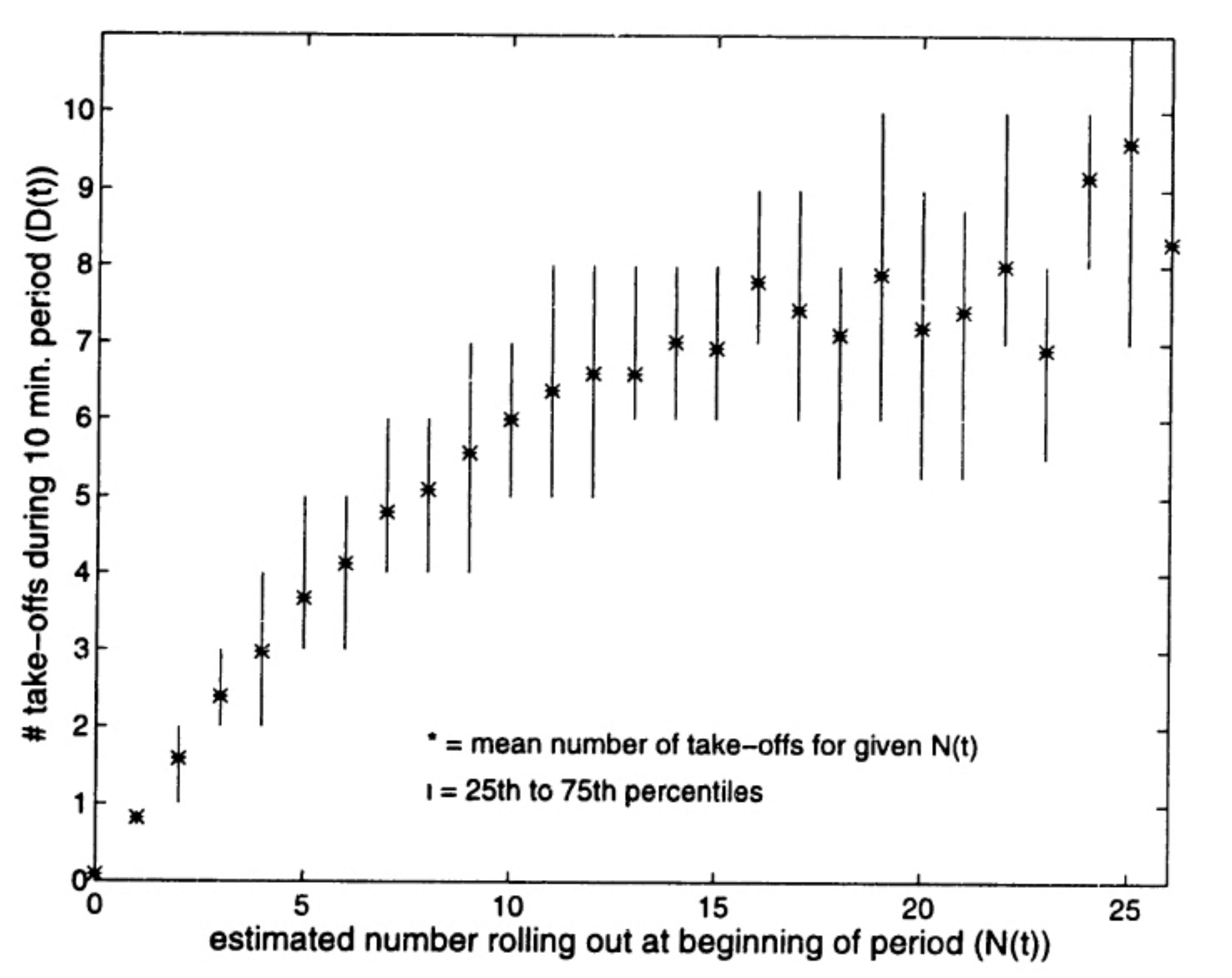}
\caption{Airport throughput as a function of surface congestion~\cite[p. 82]{Shu:95}. An asterix indicates the mean number of take-offs. Each vertical bar is the range from first to third quartile. Note how airport throughput tends to saturate when the number of aircraft taxiing-out exceed about 15. The airport is Boston Logan.}
\label{shumskypix}
\end{figure}

The fundamental observation supporting most recent research efforts is the existence of a close relationship between the number of aircraft buffered between the gate and the runway, and the runway throughput. First observed experimentally by Shumsky~\cite{Shu:95}, the runway throughput grows with the number of aircraft buffered between the gates and the taxiway; however, the throughput saturates past a given level of surface congestion, as shown in Fig.~\ref{shumskypix}.
From this observation, a number of steps followed: In~\cite{Feron1997}, Feron {\em et al.} discuss the creation of a virtual queue to control aircraft access to the taxiway system, while respecting the first-come, first-serve rule that dominates air traffic management. Pujet {\em et al.}~\cite{log99} develop a detailed queuing model of airport departure operations and introduce a simple windowing scheme similar to internet's Transmission Control Protocol (TCP) to regulate departures: Pushbacks are allowed only to the extent that the number of aircraft present on the taxiway system (the buffer), does not exceed a given threshold.
Later, Carr {\em et al.}~\cite{Carr2002} describe an approach for modeling and controlling queueing dynamic under severe flow restrictions and Idris et al. \cite{Idris2002} develop a queueing model for taxi-out time estimations. Recent developments include~\cite{Burgain2009}, which details the potential benefits of intra-airline slot-swapping inside the virtual departure queue. Finally, two notable efforts have led to field implementations of virtual queueing concepts. In~\cite{SKB:11}, Balakrishnan and her co-workers describe the experimental implementation of a congestion control scheme by means of windowing of the type proposed in~\cite{log99} and reports significant actual fuel savings and emission reductions: According to~\cite{SKB:11}, the fuel savings are of the same order of magnitude as those generated by Continuous Descent Approaches. The project CDM@CDG (see http://www.euro-cdm.org/library/airports/cdg/) has initiated the implementation of a departure manager for Charles de Gaulle Airport, whose purpose is to reduce airport surface congestion due to departing traffic. Reportedly, the virtual queueing effectively created is leveraged by airlines to perform the departure swaps studied in~\cite{Burgain2009}.

\subsection{New Surface Surveillance Information}
The gate- and spot-release control efforts described above can be easily implemented by means of existing technology. Indeed, the only information needed for implementation is already available in various forms, including ACARS~(Aircraft Communications Addressing and Reporting System), for which commercial decoding systems are now available. What motivates this paper, however, is that airports become progressively equipped with modern, digital surface surveillance technologies, such as the Airport Surface Detection Equipment, model X (ASDE-X). With such systems, accurate aircraft ground position information becomes more easily available in real-time~\cite{Eurocontrol2008,Aviation2009,Besada2004,Dimitropoulos2007}. Primarily designed for improved surface operations safety, the impact of these systems on the reduction of runway incursion incidents and conflicts has been the focus of several studies~\cite{Young2000,Singh2004,Schofield2008}.

Other studies, however, also focus on the impact of advanced surface surveillance on airport efficiency, for example
to precisely control taxiing aircraft and increase the efficiency of active runway crossings~\cite{Cheng2001}.
Early experiments show significant operational improvements enabled by airport surface surveillance technologies.
Howell et al.~\cite{Howell2004}, for example, directly measure the impact of surveillance data sharing on surface operations at Memphis International Airport and at Metropolitan Wayne County Airport. They show
that surface surveillance data made available to ground controllers directly lead to shorter taxi times.
At Memphis airport, average taxi time is reduced by 6.6 percent during Visual Approach conditions (visibility greater than five miles and ceiling greater than 5000 feet) , and by 17.5 percent during Instrument Approach conditions.
In another field study,~\cite{Howell2005} Howell et al. take advantage of a surface surveillance outage to examine its impact on airline operations.
They measure changes in taxi-out times, queue lengths, and departure rates before, during, and after the outage. They find that, for similar levels
of airport surface queues, surface surveillance decreases taxi-out times.
Furthermore, recent work investigates the practical integration
of surface surveillance for aircraft arrivals in a collaborative
environment~\cite{Andersson2003,Jung2005,Hughes2006}.

\subsection{Contributions of this paper}
The studies described above leave open, however, the analytical evaluation of the impact of these improved surveillance technologies on gate- and spot-release strategies. 
With vast amounts of data about aircraft position now available in real-time, it is legitimate to wonder whether the performance of gate-release strategies can be improved using these data. Intuitively, this should be the case: A cluster of five departing aircraft near the runway threshold should prompt decisions that differ from those required if the same cluster of five departing aircraft has just left their gate. The two situations, however, are considered to be equivalent under the policies discussed in~\cite{log99,SKB:11}.
The remaining sections of this paper therefore aim at exploring the potential benefits of high-resolution information on gate-release strategies: Section~\ref{principles} describes a modeling approach of busy airports by means of finite-state Markov Decision Processes. Section~\ref{quantitative} then discusses optimal gate- and spot-release strategies and discusses the efficiency gains that may be expected from using high-resolution surveillance systems.

\section{Modeling busy airports by means of Markov Decision Processes}
\label{principles}

To clarify the impact of added surface information on gate- and spot-release strategies, we study feedback control laws under various information scenarios. We define surface information in terms of aircraft ground position, ramp access to the taxiway system, and runway queue length. 
A stochastic model of taxi departure operations is developed by means of
Markov Decision Processes (MDP) and Partially Observable Markov Decision Processes (POMDPs). 

\subsection{Airport modeling} 
\label{s:1rModel}
Airport operations are modeled as a Markov Decision Process.
The proposed stochastic model emulates departure surface operations when:
 \begin{itemize}
 \item Exact aircraft positions are available,
 \item Aircraft trajectories are subject to uncertainties.
 \end{itemize}
Markov Decision Processes are attractive because numerical procedures are well-identified to compute optimal control policies, based on linear programming. This is unlike the models discussed in~\cite{log99}, whose resolution is finer, but which can be used only to simulate elementary control laws such as windowing schemes.
%
%
%
%
The airport surface is discretized by representing it as a finite number of ``boxes'', within which aircraft may be found.
Thus, the number of aircraft locations is finite, and depends upon a spatial sampling of the taxiway system. At each time step, aircraft may move to the next available spatial sample or stay in place. 

\subsubsection{Model Description}
When a clearance is issued by ground controllers, an aircraft enters the movement area at a given entry point corresponding to the terminal the aircraft originates from. The movement area of the taxiway-system is modeled, in this section, as a single taxiway. Aircraft motion along the taxiway is described by state transition maps, that describe the probabilities for aircraft to move forward or stay in place.
When aircraft arrive at the runway threshold, they enter a limited capacity buffer directly servicing the runway, and the aircraft order is maintained on the taxiway. 
The take-off clearance process is then simulated as a steady state stochastic process using the sum of two Bernoulli variables. This sum provides the means to calibrate, not only the average, but also the standard deviation of the take-off rate. It borrows from previous models~\cite{log99}. The uncertainty related to the take-off time illustrates the limited prediction capabilities that agents issuing ground clearances have regarding the exact take-off clearance time.

\subsubsection{Surface States Coding}
\begin{figure}[ht]
\centering
\includegraphics[width=3.5in]{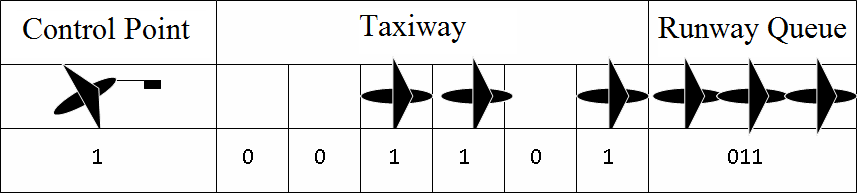}
\caption{State space model of typical airport by means of Markov decision processes. The airport features one runway, one taxiway and two access control points (spots). 
More sophisticated models could include several taxiways and runways} \label{modeling}
\end{figure}

Each state is represented as a binary vector composed of three parts: the control point, the taxiway, and the runway queue, as illustrated 
in Fig.~\ref{modeling}. The control points represent the entry points of the taxiway. When a taxi clearance is issued, 
one of the control points is switched from 0 to 1 to indicate that an aircraft was cleared to taxi toward the runway. 
The second part is the taxiway, which is directly connected to the control point. The taxiway is spatially sampled, 
with only one aircraft allowed per spatial sample. The state of the taxiway is represented by a binary vector of the same size as the taxiway. The vector's elements are set to one when the corresponding spatial sample is occupied by an aircraft, and zero otherwise. 
The runway threshold queue state
is expressed as a binary number representing the number of aircraft in the queue. For instance, if there are 3 aircraft queueing at the runway threshold, the state of the queue is given by the binary vector 011.
The entire state is then obtained by concatenating the binary vectors of the control points, the taxiway system, and the runway queue.
Finally, the overall binary vector is converted to a decimal number, 
which is its state identification number. For instance, in Fig. \ref{modeling}, the state vector 001001101011 is state 619, 
and corresponds to 3 aircraft on the taxiway and 3 aircraft in the runway queue.\\

\subsubsection{Indices and Notations}

The state space and the state index space are linked by a bijective index function.
In the rest of this paper, the notation $i$ refers to the index of a state vector.
The notation $i[s]$ refers the $s^{th}$ component of the state vector $i \in S$.

\subsubsection{Model Parameters}

The system is entirely specified by the following parameters:

\begin{itemize}
\item $L_s$ : The taxiway length represented by one spatial sample
\item $T_s$ : The sampling time
\item $N$: The number of spatial samples
\item $m$: The probability of moving forward at the next time step
\item $c_1$ and $c_2$: The probability of receiving a take-off clearance for an aircraft at the runway threshold is determined by two Bernoulli variables with parameters $c_1$ and $c_2$
\item $B$: The maximum capacity of the runway threshold aircraft buffer
\end{itemize}

The model of departure operations is a Markov Decision Process. 
Thus, it is entirely defined by the probabilities to transition from a state $i$ to another state $j$, 
knowing that  the decision to send an aircraft on the taxiway is $k$ 
(with $k=1$ corresponding to the decision of sending the next aircraft, and $k=0$ corresponding to the decision of not sending an aircraft). 
These probabilities are the model transition probabilities, and are noted $P_{j|ik}$. 
These probabilities were evaluated from the parameters described above. 
To give some idea of the model complexity, a typical airport may yield about 220,000 non-trivial transition probabilities.

\subsubsection{Markov Decision Process: States and Transition Probabilities} \label{s:transitions}

The transition probabilities are generated by enumerating through all possible simultaneous sub-transitions that lead to a feasible state. Sub-transitions are defined as atomic transitions that happen during the same time step. The process by which these transition probabilities are generated is tedious and the reader is invited to refer to~\cite{Bur:10} for more details.

\subsection{Model Calibration Procedure} \label{s:calibration}
\begin{figure}[ht]
\centering
\includegraphics[scale = .3]{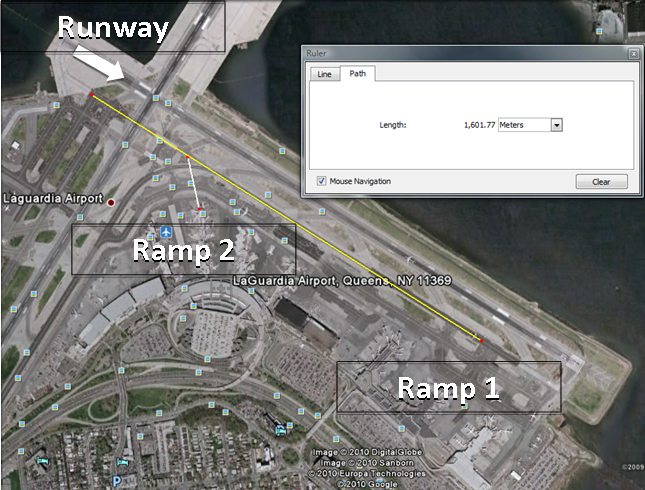}
\caption{LaGuardia airport in most common configuration.}
\label{f:LGQuickAccess}
\end{figure}

\begin{figure}[ht]
\centering
\includegraphics[scale = .3]{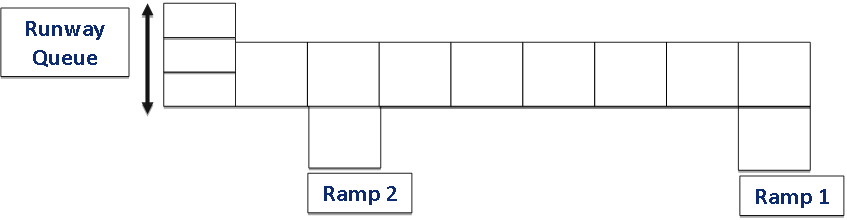}
\caption{A model of LaGuardia as Markov Decision Process - two access ramps capture the entire terminal structure}
\label{f:2ndModel}
\end{figure}

The calibration of the model is based on the analysis of the selected ASPM data, as well as 
direct observations of airport satellite pictures. The following quantities are defined: $L_s$ corresponds to observations 
of physical distances between taxiing aircraft. $T_s$ is defined as the shortest characteristic time of the different phenomena
 captured by the model. The variables $N$, the number of spatial samples of the taxiway, and $m$, the probability of moving forward 
 when unencumbered, are calibrated using taxi statistics derived from ASPM data. 
 Finally, $c_1$, $c_2$, which define the take-off probabilities, and $B$, the runway buffer size, are calibrated using take-off 
 statistics coupled with estimates of the number of taxiing aircraft. The calibration procedure is described as it is applied to New-York La Guardia airport, shown in Fig.~\ref{f:LGQuickAccess}.
Owing to the presence of two main terminals, the airport is represented using the Markov Decision Process illustrated in Fig~\ref{f:2ndModel}.
The following quantities are identified:
\subsubsection{Sampling Time}
 The temporal resolution of the ASPM data is one minute. Our model sampling frequency was set to match the sampling rate 
 of the data against which it is calibrated and $T_s$ is set to one minute.

\subsubsection{Departure Capacity}
 Heavy traffic surface operations are used to evaluate the departure capacity and 
 calibrate the take-off clearance variables $c_1$ and $c_2$. Heavy traffic corresponds to the number of aircraft for which the average 
  number of take-off per minute saturates. In the case of LaGuardia Airport, heavy traffic is achieved when 14 or more aircraft are 
  taxiing toward the runway.
  Data show that the airport throughput rate has a mean of 0.605 aircraft per minute and a standard deviation of 0.578 aircraft
per minute when the taxiway system is saturated. The take-off clearances are modeled using the sum of two Bernoulli variables $c_1$ and $c_2$ 
of parameter 0.5140, and 0.0929, respectively (variables following a Bernoulli distribution of parameter $p$ equal to 1 with success probability
$p$ and 0 with failure probability $1-p$).  The sum of the two random variables is evaluated at every minute and determines how many aircraft 
take off. 
The value of these two parameters was determined by solving the following system of equations:
\begin{eqnarray} \label{e:c1c2equation}
  \mbox{Average} = c_1 + c_2 = 0.605
  \end{eqnarray}
  \begin{eqnarray}
\mbox{Std Deviation} = \sqrt{c_1\cdot (1-c_1) + c_2\cdot (1-c_2)} =0.578
\end{eqnarray}

\subsubsection{Taxiways}

Once the departure rate variables are calibrated, the taxiway variables $N$ and $m$ are calibrated to reproduce light-traffic unimpeded taxi-time average, and standard deviation, for aircraft pushing back from each ramp area. The standard deviation and average of light-traffic taxi times were evaluated using the ASPM database. The taxi-out time is defined as the time between push-back and wheels-off and includes pushback, taxi, and waiting for take-off clearance times. 
Therefore average taxi times were computed by subtracting average pushback times and average take-off clearance times from average unimpeded taxi-out times. Likewise, taxi-time variances were computed by subtracting pushback time variances and take-off clearance time variance from unimpeded taxi-out time variances.
\begin{itemize}
\item Unimpeded taxi-out times: These taxi-out times were computed by considering taxi-out times when surrounding traffic is low. For Ramp 1, the average is 13.56 minutes and the standard deviation is 2.00 minutes.

\item Pushbacks:
Average duration of pushback was evaluated by Delcaire and Feron \cite{Delcaire1997} at 2 minutes. Based on the data collected in their report, it is fair to estimate the standard deviation of pushback duration at 80 seconds, or 1.33 minutes.

\item Take-off clearance:
Taxi-out times include waiting times for take-off clearance at the runway threshold. However, the 
model calibration should not include the variation caused by this waiting time. In this model, the average waiting time for one aircraft at the runway threshold before clearance is $1/0.605 = 1.65$ minutes, and the standard deviation $1.04$ minutes.

\item Taxi time:
According to the above discussion, the taxi time from ramp 1 has a standard deviation of $\sqrt{2.00^{2}-1.04^{2}-1.33^2}=1.07$ minutes 
and an average of $13.56-1.65-2 = 9.91$ minutes. A similar process for Ramp 2 yields an average taxi-out time from Ramp 2 equal to 6.4 minutes.

The probability $m$  of moving forward, and the number of steps $N$ from each ramp to the runway threshold, were calibrated to match the average and standard deviation of taxi times in light traffic under nominal conditions. For Ramp 1, $N$ and $m$ solve the following system of equations.
\begin{eqnarray}
\mbox{Average} = \frac{N}{m}\cdot T_s  = 9.91 \mbox{ minutes}
\end{eqnarray}
$$ \mbox{Standard Deviation} = $$
 \begin{eqnarray}
\frac{N}{m}\cdot \sqrt{\frac{1-m}{N}} \cdot T_s  = 1.07 \mbox{ minutes}
  \end{eqnarray}

Which gives,
$$N = 8.88 \approx 9 \mbox{ steps}$$
 $$m = 0.90 \approx \frac{9}{\mbox{Average}} = 0.9084.$$
For Ramp 2, $N=3$ and $m$ remains the same.
\item Calibrating the runway buffer capacity $B$:
The aircraft buffer  at the runway threshold simulates aircraft that queue closely to each other in order to ensure a high utilization rate. 
The buffer capacity must be as small as possible to limit the size of the state space over which optimal policies are computed.
However, the buffer capacity needs to be large enough to allow ground controllers to absorb uncertainties in take-off clearance time and taxi time. 
The standard deviation yielded by the sum of these two times for a single aircraft is $\sqrt{1.07^{2}+1.04^{2}}=1.49$ minutes.

The buffer was calibrated to be able, when fully loaded, to supply aircraft for a time close to 3 times this standard deviation, i.e. $4.47$ minutes. 
Thus,  the buffer size was approximated to provide enough aircraft to cover at least $4.47$ minutes, which is $4.47/0.605 = 7.39 \approx 7$ take-ff 
clearances. The capacity was set to 7 aircraft and the buffer was coded using 3 bits, as illustrated in Fig. \ref{modeling}.

\item Calibrating $L_s$:
A 200-meter separation between taxiing aircraft was suggested in previous work on taxi operations \cite{Balakrishnan2007,Visser2003}. Hence, that number was adopted here as well.\\
\end{itemize}

The calibration values for the system parameters are summarized in Table \ref{table:calibration}.

\begin{table}[h]
  \caption{Calibration values}
  \begin{center}
  \begin{tabular}{c|c}
  \hline
  \hline
  \textbf{Calibration Variables} & \textbf{Values}\\
  \hline
  $L_{s}$ & 200 meters\\
  $T_{s}$ & 60 seconds\\
  $N$ & 9 (Ramp 1)\\
  & 3 (Ramp 2)\\
  $m$ & 0.9084\\
  $c_1$ & 0.5140\\
  $c_2$ & 0.0929\\
  $B$ & 7\\
  \hline
  \hline
\end{tabular}
\end{center}
\label{table:calibration}
\end{table}

\subsubsection{Model Validation}
\begin{figure}[ht]
\centering
\includegraphics[width=3.5in]{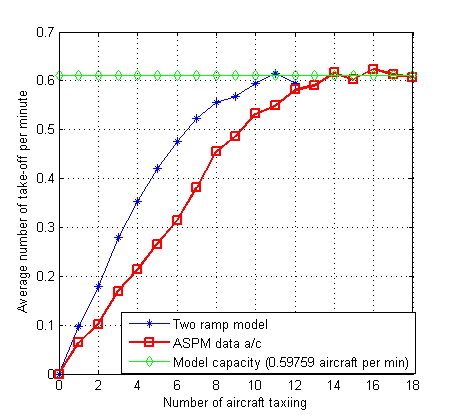}
\caption{LaGuardia throughput as a function of the number of taxiing aircraft, from the two ramp model and ASPM data. The ASPM data reflects all departure operations on the ground starting at pushback.}
\label{loadGraph21}
\end{figure}

\begin{figure}[ht]
\centering
\includegraphics[width=3.5in]{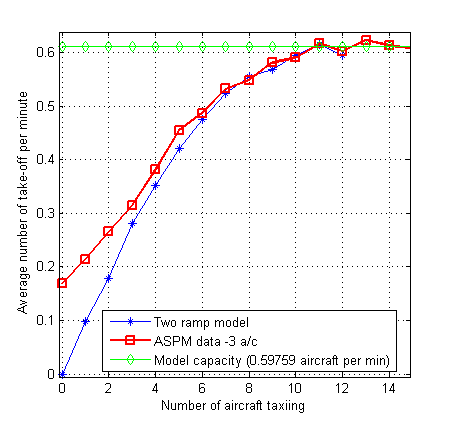}
\caption{LaGuardia throughput as a function of the number of taxiing aircraft, from the two ramp model and ASPM data. The ASPM curve is shifted by 3 aircraft to isolate taxiway operations starting at ramp exit control points, for utilization rates above 30\%.}
\label{loadGraph22}
\end{figure}

Using ASPM data, La Guardia airport average throughput rate is expressed as a function of the number of taxiing aircraft. The graph provided in
  Fig. \ref{loadGraph21} shows the airport throughput as a function of the number of taxiing aircraft, and yields the average take-off rate.
Fig. \ref{loadGraph21}  also shows the throughput as a function of the number of taxiing aircraft for the stochastic model.
The model behaves similarly to the airport, and faithfully reproduces the queueing and stochastic nature of departure operations. 
When the number of taxiing aircraft reaches 11, the model saturates, and yields a maximum take-off rate distribution averaging 0.598 
aircraft per minute, with a standard deviation of 0.585 aircraft per minute. These are similar to the average (0.605) and the standard 
deviation (0.578) of the observed take-off rate at LaGuardia, when the taxiway is saturated by departing aircraft. The saturation level of the 
model take-off rate is reached at a lower number of taxiing aircraft than for the ASPM data because the model accounts for operations
 on the taxiway only from the ramp control points. By contrast, the ASPM data includes all aircraft on the ground starting at pushback.
  The ASPM data does not provide aircraft position, therefore it is not possible to distinguish aircraft still pushing back at the ramp 
  from aircraft which are at the ramp exit control points. To isolate taxiway operations starting at the control points from the rest of
   the ramp operations in the ASPM data, the ASPM curve has been shifted to match the saturation level of both curves. For runway utilization
    rates above 30\% of interest in this paper, the shift efficiently isolates taxiway operations starting at the control points in the ASPM data, as illustrated 
    in Fig. \ref{loadGraph22}. Note that the two-ramp model performs better than the one-ramp model.  

\section{Quantitative impact of full-state information: Optimal control of airports represented as MDPs}
\label{quantitative}

To understand and valuate the impact of aircraft position information on departure operations, an approach based on the optimization of Markov Decision Processes (MDP) and Partially Observable Markov Decision Processes (POMDP) was developed. This approach is applied to two distinct state-based policies and one benchmark policy:
\begin{itemize}
\item The first policy, named ``optimal full state feedback'' assumes that the state of the surface is fully available.
\item The second policy, named ``estimated state feedback'' assumes that the only part of the state of the surface is known.
\item The benchmark policy, named ``threshold policy'' is that used in prior analytical and experimental works~\cite{log99,SKB:11}.
\end{itemize}

\subsection{Approach} \label{theappraoch}

The objective of this approach is to evaluate how the level of information available on aircraft position affects potential taxi-time reductions, for a given rate of runway utilization, and within a collaborative framework enabling the fine tuning of taxi clearances, when aircraft exit the ramp area.

\subsubsection{Assumptions}
Ground controllers operate as optimally as allowed by existing technology: they know the behavior of the system, and given the level of information available, they understand what the best policy is. Their goal is to maximize the departure runway utilization rate, while controlling aircraft to minimize taxi times. It is assumed that there is enough departure demand for FAA ground controllers to always have an aircraft waiting to be cleared for taxi at both ramps, since this corresponds to peak demand times. The aircraft is either cleared for push-back, if it pushes directly on the movement area, or cleared for taxi, if it has already pushed back on the ramp area, and is waiting at a control point to enter the movement area.

\subsubsection{Optimal pushback policies}
 Each state has a cost, an optimal clearance policy is the set of decisions that minimizes the expected averaged cost over an infinite time horizon.

\subsubsection{Trade-offs and cost structure}
For each time instant $i$, each state $i$ is given a cost $C_{i}$ that reflects its desireability. This cost is a weighted sum of the number of taxiing aircraft  $N_{ac}(i)$ and a cost attributed to the non-utilization of the runway $\delta_{r}(i)$ multiplied by a constant $\beta$. The variable $\delta_{r}(i)$ is equal to 1 if there is no aircraft in the runway buffer and to 0 if there is at least 1 aircraft.
 For every state $i$, the cost $C_{i}$ attributed to that state is given by
 \begin{eqnarray}
  \label{costi}
C_{i} = N_{ac}(i) + \beta \cdot \delta_{r}(i).
\end{eqnarray}

As $\beta$ increases, the optimal policy favors maximizing the runway utilization rate over minimizing the number of taxiing aircraft. $\beta$ and $C_{i}$ are expressed in number of aircraft per minute. $\beta$ is the ratio of the cost of non-utilization of the runway for one minute over the cost of having one aircraft on the taxiway for one minute.\\
For each value of $\beta$, the corresponding optimal policy is Pareto optimal and captures the trade-off between minimizing taxi time
 and maximizing runway utilization rate. 
 
\subsubsection{Fairness considerations when multiple ramps are present}
It is assumed that each terminal (ramp) has aircraft ready to enter the taxiway system, and that they must be serviced fairly. Two mechanisms have been introduced for that purpose: In {\em ramp alternation}, the policy must service each ramp once at a time. An additional state is introduced in the Markov Decision process to reflect this. In {\em statistical fairness}, a constraint is introduced to constrain each ramp to be serviced an equal number of times {\em on average}.

\subsection{Information valuation}
The metrics used to value information are runway utilization and number of taxiing aircraft. 
The value of added information is computed as the improvement in closed-loop system performance generated by this added information.

\subsubsection{Full State Feedback and optimal policies}\label{s:LP}

Under full state feedback, the agent controlling the clearances can fully observe state of the surface state.
The optimal decision $k$ is a function of the observed state $i$. Given the cost structure and the representation of the airport taxi-out process as a Markov Decision Process, it is possible to use linear optimization techniques to find the steady-state optimal decision policy $\Pi$ that minimizes
the expected cost per time step~\cite{Hiller2001}. If $i(t)$ is the state at time $t$, then
 \begin{eqnarray}
 \mbox{Expected Cost} = \displaystyle\lim_{n \to \infty} E \left( \frac{1}{n} \cdot \displaystyle\sum_{t=0}^n C_{i(t)} \right).
\end{eqnarray}
To detail the optimal control approach, we use the following notations:
\begin{itemize}
\item Let $\iota$ be the state at time n.
\item Let $\eta$  be the state at time n+1.
\item Let $\kappa$ be the decision variable value at time n.
\item Let $y_{ik} = P(\iota=i,\kappa=k)$ be the probability of being in state $i$ and taking decision $k$. The optimal decision $k$ is given by the optimal policy: $k = \Pi(i)$.
\item Let $p_{j|ik} = P(\eta=j|\iota=i,\kappa=k)$ be the probability of having the next state $j$ knowing the current state is $i$ and the decision chosen is $k$.
\\item In addition, a state is added that describes whether the next pushback originates from ramp 1 or ramp 2 
\end{itemize}

For a steady state process with $M+1$ states and $K$ decisions, the expected cost per time step is \cite{Hiller2001}
   \begin{eqnarray}\label{e:costEquation}
   \displaystyle\lim_{n \to \infty} E(\frac{1}{n} \cdot \displaystyle\sum_{t=0}^n C_{i(t)}) = \displaystyle\sum_{i=0}^M \displaystyle\sum_{k=1}^K C_{ik}\cdot y_{ik}.
    \end{eqnarray}

Consequently, the cost function for this linear optimization is
 \begin{eqnarray}\label{e:costEquationFinal}
\mbox{Minimize } Z = \displaystyle\sum_{i=0}^M \displaystyle\sum_{k=1}^K
C_{ik}\cdot y_{ik}.
 \end{eqnarray}

Subject to:
\begin{enumerate}
\item Constraints on state-decision probability variables:
\begin{eqnarray} \label{e:stchastiVariableConstraints}
\displaystyle\sum_{i=0}^M \displaystyle\sum_{k=1}^K y_{ik} = 1  \end{eqnarray}
 \begin{eqnarray} \label{e:stochasticVariableConstraints}
  y_{ik} \geq  0, \mbox{for } i = 0..M; k = 1..K \end{eqnarray}

\item Constraints governing state transitions:
 \begin{eqnarray}\label{e:stateTransitionEq}
 \displaystyle\sum_{k=1}^K y_{jk} - \displaystyle\sum_{i=0}^M \displaystyle\sum_{k=1}^K y_{ik}\cdot p_{j|ik} = 0,
\end{eqnarray} \\
\mbox{for } $j = 0..M$; $k =1..K$

\end{enumerate}

Once the optimal set of steady state probabilities of being in state $i$ and taking decision $k$, $y_{ik}$, is evaluated and the corresponding 
optimal pushback policy is given by

\begin{equation}
\label{e:policyEvaluation}
\mbox{if } y_{ik} > 0 \mbox{ then } \Pi_{i} = 1
\mbox{ else if } y_{ik} = 0 \mbox{ then } \Pi_{i} = 0
\end{equation}

\subsubsection{Partial information: Estimated State Feedback} \label{partial}

In this scenario, the agent has access to the number of taxiing aircraft, and he knows whether or not it is physically possible to clear an 
aircraft (there may be another aircraft in the way). In real-life situations, this partial information is always available because ramp 
controllers keep track of the number of aircraft that have pushed back, and the number of aircraft that have taken-off.
 In addition, communication technologies ensure simple Input-Ouput information is easily available. For instance, Aircraft Communications 
 Addressing and Reporting System (ACARS) is a digital datalink system for transmission of short relatively simple messages between aircraft 
 and ground stations via radio or satellite and provides aircraft take-off times.
  Under limited aircraft position information, the system becomes a Partially Observable Decision Process (POMDP). There exists several methods
   to solve POMDPs optimally \cite{Cassandra1996,Cassandra1997,Sondik1971,Kaelbling1998}.
  These methods are computationally very demanding for a finite time horizon, and not appropriate for an infinite time horizon. Indeed, finite-horizon 
  POMDPs are PSPACE-complete \cite{Papadimitriou1987} and infinite-horizon POMDPs are undecidable \cite{Madani1999}.

\paragraph{Most Likely State}

\begin{figure}[ht]
\centering
\includegraphics[scale = .4]{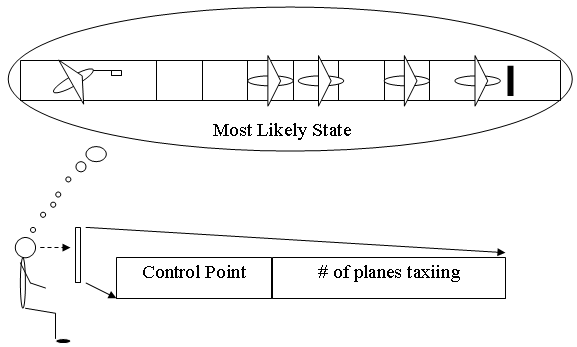}
\caption{Estimation of the taxiway system state by a decision maker}
\label{MostLikelyState}
\end{figure}

For these reasons, methods applicable to an infinite time horizon and computationally more tractable were considered \cite{Cassandra1996}. 
The Markov process that is modeled for LaGuardia includes more than 220,000 transitions with non-zero probabilities. 
Heuristic methods are computationally faster, and better suited to determine effective control laws for this POMDP.
 Among these, the Most Likely State (MLS) algorithm was selected because it is applicable to an infinite time horizon and 
 compares favorably with other effective heuristic algorithms \cite{Cassandra1996}. 
 Moreover, its steps resemble the behavior of a decision maker under uncertainty. 
 Indeed, this heuristic control strategy consists of estimating the most likely current state, and choosing the corresponding optimal decision, using the optimal decision policy evaluated in the full state feedback case. Fig. \ref{MostLikelyState} illustrates the information available to the decision maker.\\

The variables used in the MLS algorithm are defined as follow:
\begin{itemize}
\item Let $\Theta$ designate the index of the current observation. The current observation is the number of 
taxiing aircraft and wether or not it is physically possible to clear an additional aircraft for pushback or taxi.
\item Let $b_i$ be the probability of being in state $i$ for all states $i$ of the state space. $b$ is the belief state vector.
\item Let $p_{o|j} = P(\Theta=o|\eta=j)$ be the probability of observing $o$, knowing the current state is $j$, for all observations $o$ in the observation space, and for all states $j$ in the state space.
\end{itemize}
The updater function takes the previous belief state $b$, the current observation $o$, the previous decision $k$, and returns the current belief state vector  $b'$.\\ The following equation is derived from Bayes' rules \cite{Littman1994}:
\begin{equation}
  \label{e:updater}
 b'_{j} = \frac{p_{o|j} \displaystyle\sum_{i=0}^M p_{j|ik} \cdot b_{i} }{ \displaystyle\sum_{j=0}^M p_{o|j} \displaystyle\sum_{i=0}^M p_{j|ik} \cdot b_{i}}.
 \end{equation}

\begin{figure}[ht]
\centering
\includegraphics[width=3.5in]{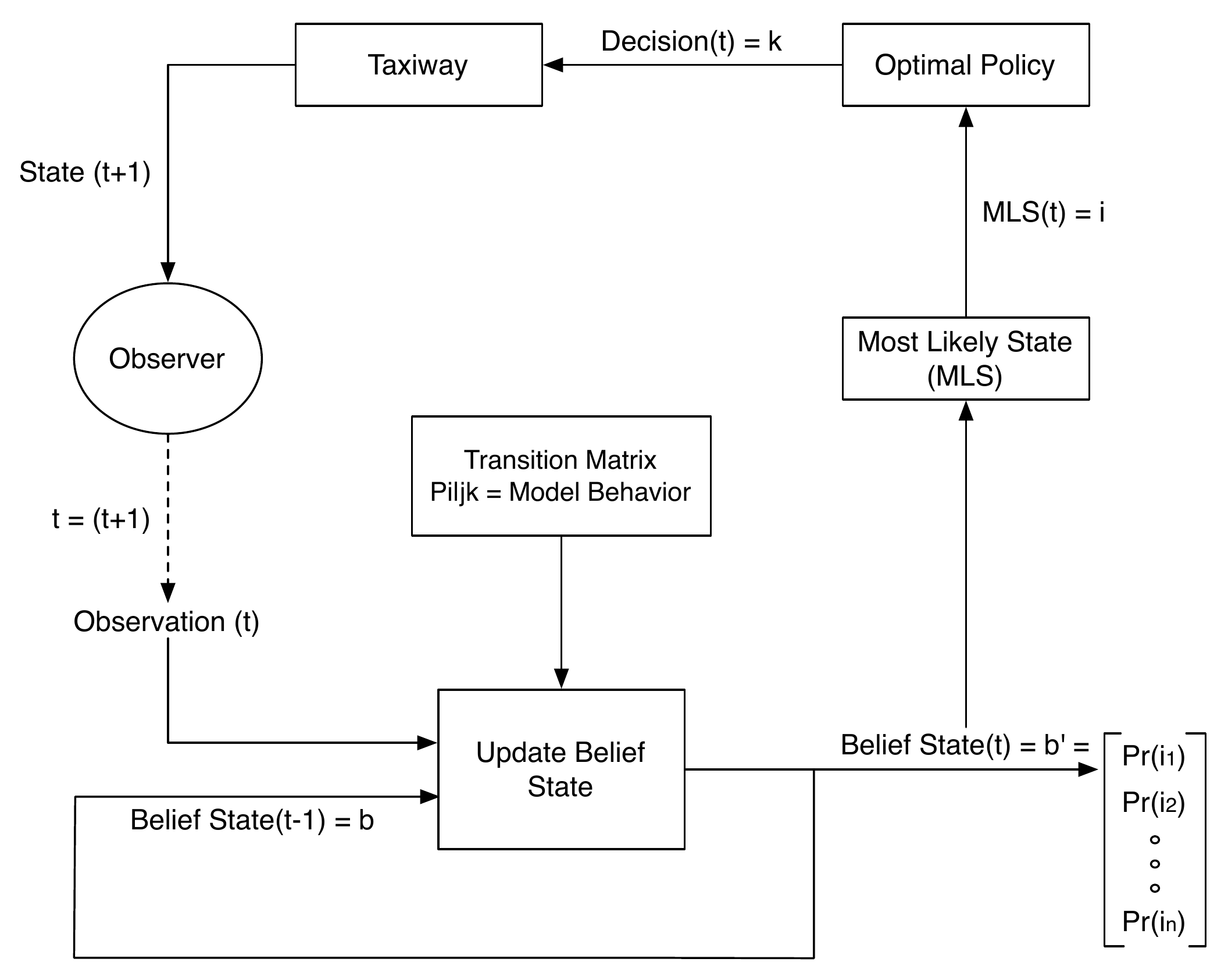}
\caption{Heuristic control of taxi clearance decisions based on partial observation}
\label{estimator}
\end{figure}

Fig. \ref{estimator} details the heuristic control of taxi clearance decisions based on partial observations.

\paragraph{Observation probability matrix}

This paragraph details how observations are defined and how  the probability matrix $p_{o|j}$ is evaluated.
The information contained in an observation is given by the probability matrix $p_{o|j}$. This probability is key to evaluate the probability of having every state $j$ given a specific observation $o$ and previous belief $b$. Eq. (\ref{e:updater}) explains how observations of the surface are incorporated into the decision process during the update of the belief state.

Let $O$ be the observation space and $c$ be the total number of components, or piece of information, included in each observation $o \in O$. 
Then $O$ is a subset of $\Re^c$. In this scenario there are two pieces of information, $c = 2$, the number of taxiing aircraft $N_{ac}$, 
and whether or not it is physically possible to clear an aircraft using ($RampFree$), a binary variable.
An observation is a vector defined by

\begin{equation} \label{e:oVect0}
o = \begin{bmatrix} N_{ac} & RampFree \end{bmatrix}.
\end{equation}

The algorithm generating the observation probability matrix uses an injective function,
 which attributes a unique observation index $n(o)$ to every observation $o$. 
 The injective function converts the observation vector with 2 components into a binary vector of $[roundup(log2(max(Nm_{ac})))+roundup(log2(max(RampFree)))]$ bits to then reconvert it back to its decimal value, as illustrated in Eq. (\ref{e:obsInjFunct}).

\begin{equation}
\label{e:obsInjFunct}
n(o) = bin2dec(\begin{bmatrix} dec2bin(N_{ac}) & dec2bin(RampFree)\end{bmatrix})
\end{equation}

For any state $j$, there exists only one information that can be observed $o(j)$, consequently for a system with $N$ possible observations and $M$ states, observation probabilities are zeros and ones, i.e. $\forall (o_n,j) \in \{1..N\}\times \{1..M\}$, $p_{o_n|j} \in {0,1}$. Eq. (\ref{e:obsPoj}) shows how the $p_{o|j}$ matrix is evaluated.

\begin{equation}\label{e:obsPoj}
p_{o_n|j} =
\begin{cases}
1 \text{ if $o_n = n(o(j))$}, \\
0 \text{ if $o_n \neq n(o(j))$}.
\end{cases}
\end{equation}

\subsubsection{Threshold policy}
A threshold policy is a pushback control law, which relies solely on the current number of taxiing aircraft to make a push-back decision and is described in~\cite{log99,SKB:11}.
 This simple control law computes the number of taxiing aircraft  $N(i)$ for state $i$ and compares it to a given threshold value $Th$ \cite{log99}.
  If the number of aircraft is greater than the threshold, no pushback clearance is issued, and $k=0$. On the other hand, if that number is smaller 
  than the threshold, a pushback clearance is issued, and $k=1$. This is summarized by the following Eq. (\ref{e:theq1}),


\begin{equation} \label{e:theq1}
k =
\begin{cases}
1 \text{  if $N(i) > Th$},\\
0 \text{  if $N(i) \leq Th$}.
\end{cases}
\end{equation}

Note that the threshold policy can be evaluated analytically since the corresponding closed-loop system is a Markov chain. When multiple ramps are present, the threshold policy is required to alternate evenly among the ramps.

\subsection{Optimal policies against benchmark policy: New York La Guardia airport}

\begin{figure}[ht]\hspace{-5mm}
\includegraphics[scale = 0.5]{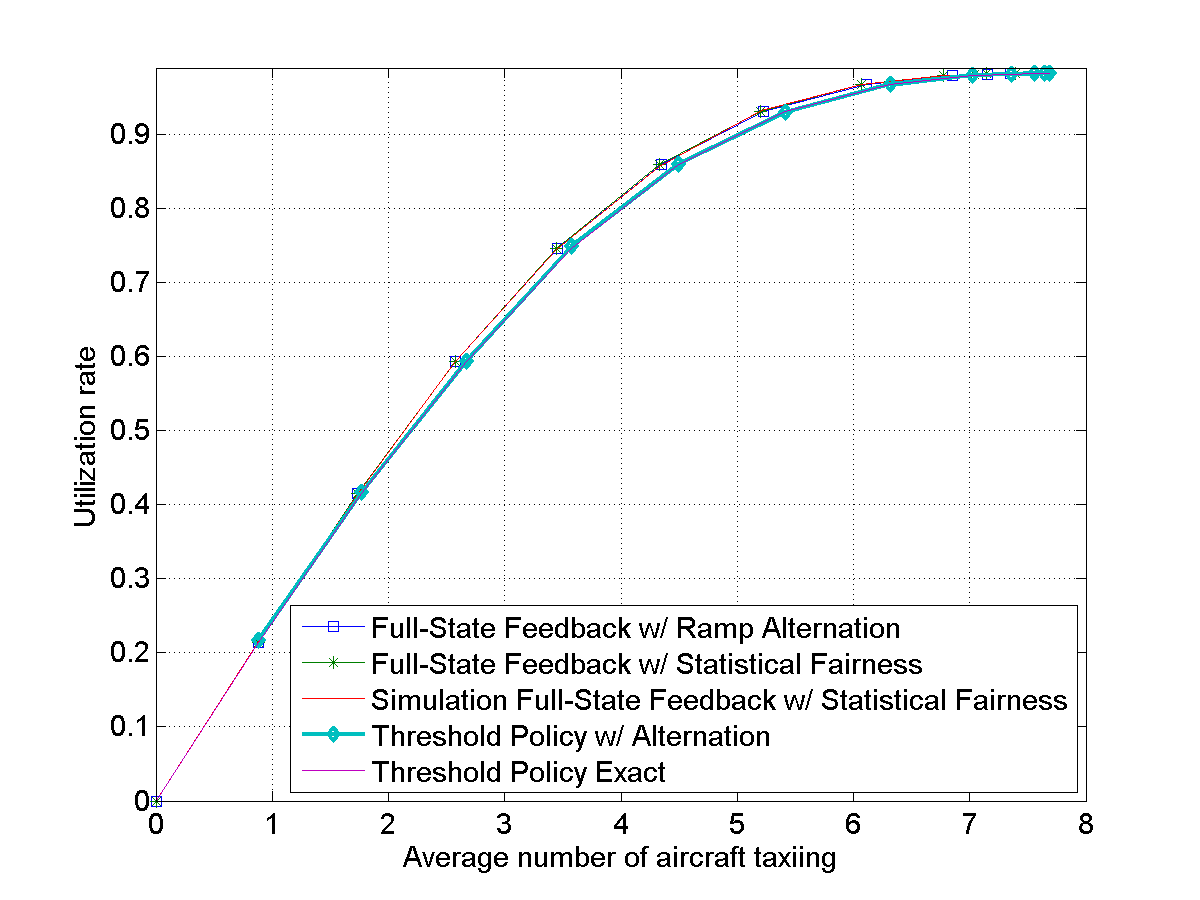}
\caption{Runway utilization rate as a function of the average number of taxiing aircraft at LaGuardia Airport}
\label{f:CXutilVsTax}
\end{figure}

\begin{figure}[ht]
\hspace{-5mm}
\includegraphics[scale = 0.5]{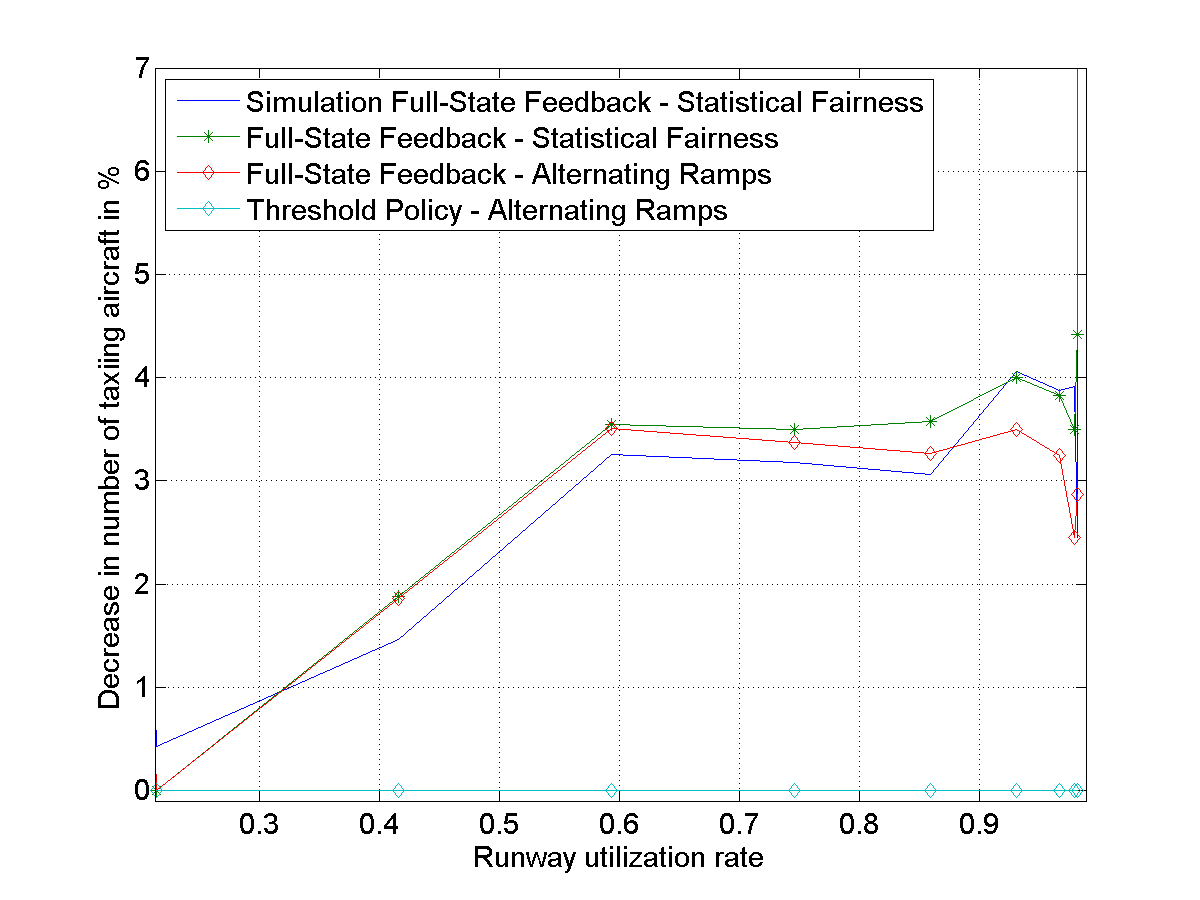}
\caption{Reduction in percent of the average number of taxiing aircraft as a function of the utilization rate,
when compared with a threshold policy which alternates between ramp one and ramp two.}
\label{f:CximprovVsUtil}
\end{figure}

Fig. \ref{f:CXutilVsTax} illustrates the utilization rate of the LaGuardia Airport two ramp model 
as a function of the average number of taxiing aircraft. 
Fig. \ref{f:CximprovVsUtil} shows the reduction in percent of the average number of taxiing aircraft for optimal policies, 
as a function of the utilization rate,
when compared with a threshold policy which alternates between ramp one and ramp two.

When the number of taxiing aircraft is limited to one aircraft by the threshold policy, the difference of performance between the full-state
 feedback policy and the threshold policy is inexistent, as illustrated by Fig. \ref{f:CximprovVsUtil}. 
 This confirms the intuition that there is no benefit in knowing the exact  position of aircraft 
 when there can be no conflict between aircraft on the taxiway.

When the threshold for the number of taxiing aircraft is increased to two and and then three, 
the threshold policy starts yielding a lower utilization rate for the same number of taxiing aircraft than the full-state feedback policy, 
as shown in Fig. \ref{f:CXutilVsTax} and \ref{f:CximprovVsUtil}. Indeed, the threshold policy releases aircraft blindly, based on the number 
of taxiing aircraft. Consequently aircraft have a higher probability of conflicting with each other on the taxiway. 
The optimal full-state feedback policy performs better because it manages the release of aircraft using the exact position of the other aircraft already taxiing.

Among the two fairness rules described above, the policy based on statistical fairness yields the best results, 
as shown in Fig. \ref{f:CximprovVsUtil}. However, it performs close to the policy that strictly alternates between ramps. 
It is noticeable that the \emph{simulation} of the statistical fairness optimal policy, produces performances that are slightly
 worse than those directly indicated by the optimization software output. 
 Consistent with the reliability analysis, there is a sharp divergence of reliability when the runway utilization rate exceeds 0.98.

The full state feedback policies perform consistently better, generating a smaller average number of taxiing aircraft, 
when compared with the threshold benchmark policy. This performance is consistently better over a wide array of runway utilization rates,
 which correspond not only to intermediate runway capacities, but also to situations where the runway is used at maximum capacity. 
 For rates between 0.6 and 1, the reduction of the number of taxiing aircraft is consistently above 3.5 percent. 
 Most interestingly, as the runway utilization rate increases from 0.92 to 0.96, the savings reach 4 percent.

\subsection{Extensions}
The methodology presented here has been extended to other airports with simple runway/taxiway structures. Considering Seattle-Tacoma airport (SEA),
a similar modeling approach, followed by optimal control policies, has allowed us to evaluate the savings (in terms of reduced emissions, or number of aircraft on the airport surface during busy hours) at approximately 6\% compared with threshold approaches currently under evaluation. 

Extensions of this work to large airports featuring multiple terminals, taxiways and runways are possible. However, several challenges must then be addressed. First, extensive, high-resolution datasets must be available to the user for model calibration. Unlike the ACARS data, which are sufficient for the calibration of simple airport dynamics, more extensive datasets, such as those generated by surface monitoring systems such as ASDE-X are unlikely to be as easy to manipulate. Second, the design of optimal control policies requires the solution of linear programs whose size, commensurate with the underlying state-spaces, largely exceeds existing computational capabilities. As a result, alternative design techniques may need using, such as approximate dynamic programming techniques~\cite{BeT:96}. 

\section{Conclusions}
This paper assesses the benefits of providing surface surveillance information to the ramp clearance control
process at busy airports.

Our results have shown that, within a collaborative framework allowing the creation of a virtual queue, surface surveillance information can significantly improve the control of stochastic 
departure operations on the ground. More specifically, at LaGuardia airport, controlling taxi clearances optimally using surface surveillance reduces
 the number of taxiing aircraft by 4\% when the airport functions near capacity, compared with a threshold policy which limits the number of taxiing 
 aircraft. At Seattle airport, controlling taxi clearances optimally using surface surveillance reduces the number
of taxiing aircraft by 6\% when the airport functions near capacity, compared with a threshold policy which limits the number of taxiing aircraft.
It has been observed that, in order to minimize wasteful surface conflicts and queues, the optimal full-state feedback policy relies on aircraft position information to avoid conflicts, maximize runway utilization, and balance and coordinate ramp taxi clearances. 


\section*{Author Biography}
\noindent {\bf Pierrick Burgain} is an analyst at Metron Aviation. He holds a PhD in Electrical and Computer Engineering and an MBA from the Georgia Institute of Technology. His research interests include the application of stochastic control and optimization to applications in air traffic operations.\\

\noindent {\bf Eric Feron} is Dutton/Ducoffe professor of aerospace software engineering at Georgia Tech, USA. A graduate of Ecole Polytechnique, France, and Stanford University, USA, he has also held appointments with MIT, USA and ONERA, France. His research encompasses applications of control systems, computer science and optimization theory to several problems of aerospace interest.

\end{document}